# Effect of Group Means on the Probability of Consensus

YOSHIKO ARIMA, Kyoto Gakuen University

## 1. INTRODUCTION

When making a group decision, some groups reach a consensus quickly, whereas others take more time. What factors affect the probability of consensus within a certain period has not been investigated. Using group polarization paradigm, this study examined the antecedents of intragroup disagreement and explored attitude changes in disagreement groups compared to consensus groups. A "disagreement group" was defined as a group that cannot reach a consensus within a given period. Group polarization (Moscovici & Zavalloni, 1969) is a phenomenon of exaggeration of individual attitudes through group interaction (Lamm & Myers, 1978).

### 1.1 Group Polarization

Group polarization has been studied by social comparison theorists (Sanders & Baron, 1977) and informational influence theorists (Burnstein & Vinokur, 1977). Research interest in this area has declined, partly because interest in decision-making theories has increased. Using simulation models, proponents of social-decision-scheme (SDS) theory (Davis, 1973) have suggested that polarization is mainly the product of majority rule, implying that socially shared values do not necessarily influence group polarization (Kerr, Davis, Meek, & Rissman,1975). Research based on self-categorization theory, however, has shown that social categorization can alter the direction of group polarization (Abrams, Wetherell, Cochrane, Hogg, & Turner,1990) .

These studies suggest two directions of influence, from a subordinate level to a super-ordinate level and from a super-ordinate level to a subordinate level. The opposing directions of influence in different levels would be present in the intragroup disagreement process. Under this assumption, it was hypothesized that both between-group discrepancy and within-group discrepancy would influence probability of consensus and attitude change within a disagreement group.

## 2. Method

### 2.1 Participants and Procedure

A total of 269 females participated in the experiment: (a) 93 students from secretary course at the Women's College, (b) 43 nurses from a leadership training seminar conducted by their hospital, (c) 73 homemakers from a daytime seminar for community leaders conducted by the Hyogo Prefecture Association, and (d) 60 part-time working women from an evening seminar for community leaders conducted by the Hyogo Prefecture Association.

After completion of the pre-test questionnaire, the researcher randomly assembled 56 four-person groups and 9 five-person groups. Each group was delivered two sheets of questionnaire, and instructed that they could split their decision if they could not reach a consensus within 60 minutes. Groups who reached a consensus on all 10 items were labeled *consensus groups* and those who could not were labeled *disagreement groups*. After the group decisions were recorded, the members answered a post-test questionnaire that contains the same items as the pre-test.

### 2.2 Independent and dependent variables.

The discussion topic was conservative vs. liberal attitudes toward women in the workforce. The initial mean of each group is called the "group mean", the initial mean of all the members of a sample is called the "grand mean", and initial variance of each group is called the "group variance." 'Initial mean tendency" refers to the direction of polarization. For example, on the 7-point scale (from

disagree(1) to agree(7), midpoint =4) used in this research, if a group mean is 3.5, the initial mean tendency is in the "con" direction, while if a grand mean is 4.5, that tendency is in a "pro" direction.
The dependent variables were the probability of consensus, defined as the proportion of all groups who reached a consensus, and participant shifts, defined as the sum of the 10 shift scores on post-test minus pre-test) which represents the change in attitude of the members of each group.

## 3.  Results

A factor analysis (MLC) of the 10 pre-test items produced 2 factors, each eigenvalue 3.23 (32.28%), 1.4(13.97%), respectively. Bartlett's Test of Sphericity shows significant correlation among variables ($\chi^2$= 513.85, df= 45, p<.001). Cronbach's α for the 10 items was .76. Cronbach's α for the 10 items of pre-test was .76 (ICC=.19) and that of post-test was .83 (ICC=.26). The overall pretest mean was 43.61, indicating that the initial mean tendency was in the liberal direction. The mean attitude of the low group mean was 39.61 ("con" direction, which predicts shift for minus direction), and the mean attitude of the high group mean was 47.69 ("pro" direction which predicts shift for plus direction).

### 3.1   Probability of consensus

A logistic regression analysis was conducted to predict the probability of consensus from the standardized group means, standardized group variances, interaction between the group means and variances, and subgroups (nurses, students, homemakers, and part time homemakers). These variables were entered into the logistic regression. Fig. 1 shows the results for the probability of consensus. There was a significant interaction effect between the group means and variances for the probability of consensus [$B$ = .55, Wald (1) = 3.81, $p < .05$, $\chi^2$ (1, $N$ = 65) = 4.47, $p < .05$]. The probability of consensus was lower in the low-mean groups than in the high-mean groups, and the interaction effects with the group variances only appeared in the high-mean groups.

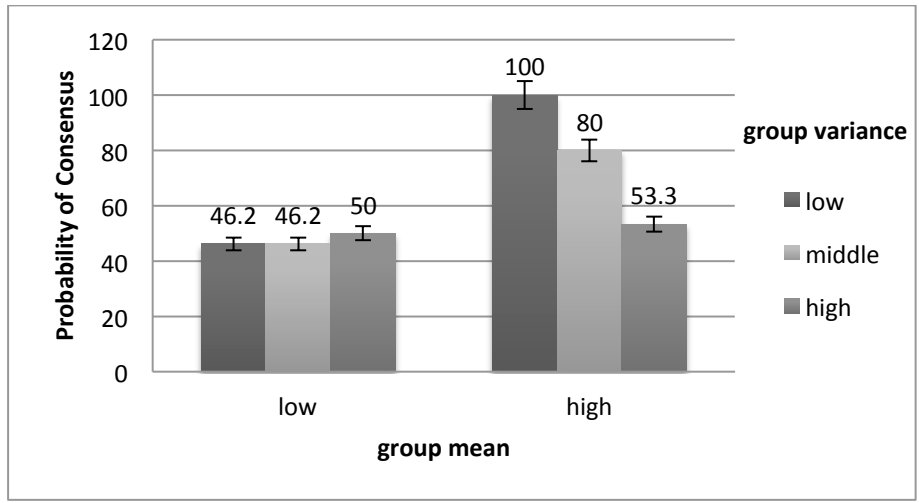

Fig. 1. The probability of consensus as a function of the group means and variances.

### 3.2   Participant's attitude change

Fig. 2 shows the participant shift scores for the consensus and disagreement groups. Regression analyses were conducted on the consensus and disagreement groups, respectively, for the standardized group means, the standardized group variances, and their interactions. The analysis for the consensus groups yielded a significant main effect of group mean on the total shift scores [$B = 0.73$, adjusted β = .64, $t_{152} = 2.99$, $p < .01$, adjusted $R^2$ = .14]. In contrast, the analysis for the disagreement groups yielded a significant interaction effect between group means and group variances on the total shift scores [$B = 0.07$, adjusted $\beta = 1.7$, $t_{99} = 2.29$, $p < .03$, adjusted $R^2 = .13$]. While group polarization was observed for the consensus groups, negative shift scores (depolarization) were found regardless of the group mean for the low-variance disagreement groups.

Fig. 3 shows two scatter diagrams comparing the group means and variances of the consensus and disagreement groups for the pre-test and post-test, respectively. These plots suggest that the within-group variances converged (decreased) from pre-test to post-test ($Ms = 12.49, 8.79$) regardless of discussion outcomes. In contrast to the within-group variance, the between-groups variance increased from pre-test to post-test ($SDs = 5.40, 7.63$).

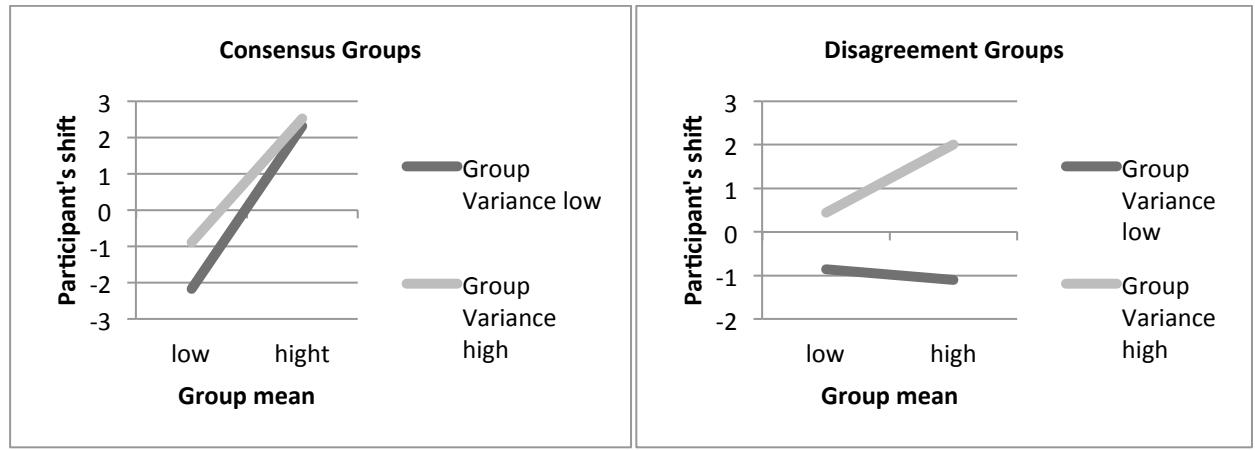

Fig. 2. Participant shift scores as a function of the consensus group and disagreement group means and variances.

4.  Conclusion

(1) In the consensus groups, groups' initial mean tendency determined the direction of attitude change. There was not an influence from super-ordinate level.

(2) The probability of consensus was determined by the relative position of group means in the means distribution. There was an influence from super-ordinate level.

(2) In the disagreement groups, the direction of attitude change was controlled by the within-group variance. If the group's variance was low, the direction was reversed from the grand mean's tendency.

(4) While the within-group variances converging, the between-group variance increased after the group discussions.

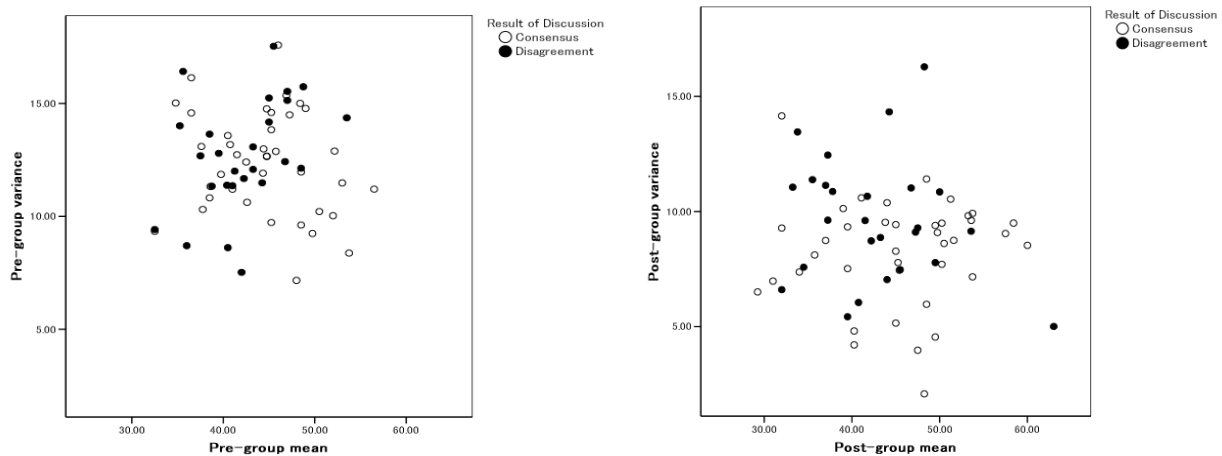

Fig. 3. Scatter plots of the pre-test and post-test means and variances for the consensus and disagreement groups.